\documentclass[final,5p,times,twocolumn]{elsarticle}

\usepackage[utf8]{inputenc}
\usepackage[T1]{fontenc}
\usepackage{xcolor}

\usepackage{amssymb}
\usepackage{amsmath,amsfonts}%
\usepackage{booktabs}
\usepackage{multirow}

\biboptions{sort&compress}

\journal{Infrared Physics \& Technology}

\begin{document}

\begin{frontmatter}


\title{Broadband Two-Dimensional Far-Field Beam Profiling of Commercial CW Terahertz Photomixers from 0.2 to 1.5 THz}

\author[1]{Mathias Hedegaard Kristensen\corref{cor1}}
\cortext[cor1]{Corresponding author. Presently at University of Bordeaux, CNRS, LOMA UMR 5798, 33405 Talence, France.}
\ead{mhkr@icloud.com}
\affiliation[1]{organization={Department of Materials and Production, Aalborg University},
            city={Aalborg East},
            postcode={DK-9220}, 
            country={Denmark}}

\author[1]{Esben Skovsen}
\ead{es@mp.aau.dk}

\begin{abstract}
We present a systematic far-field characterization of four commercial PIN diode terahertz photomixers over the 0.2–1.5\,THz frequency range using a two-dimensional (2D) raster scanning method.
The emission pattern of each transmitter, equipped with an integrated hyper-hemispherical silicon lens, was characterized using a broadband Schottky diode receiver over a 35$\times$35\,mm$^2$ grid.
The results reveal consistent frequency-dependent beam divergence and the emergence of distinct Airy diffraction patterns at intermediate frequencies, attributed to lens-induced aperture effects.
In addition to qualitative mapping, we extract quantitative metrics -- including divergence slope, beam asymmetry, ellipticity, and centroid variability -- that enable objective comparison of beam profiles across devices and frequencies.
This comprehensive mapping underscores the importance of full 2D beam profiling for understanding THz propagation and offers insights into lens design and photomixer packaging for optimized system performance.
\end{abstract}




\end{frontmatter}


\section{Introduction}
Photomixers are a well-established technology for generating continuous-wave (CW) terahertz (THz) radiation, offering broad frequency tunability and room-temperature operation. These attributes make them highly attractive for applications in imaging, spectroscopy, and high-speed wireless communications. However, a critical yet often underexplored aspect of photomixer performance is the spatial emission pattern of the THz beam, which directly affects coupling efficiency, optical alignment, and the integration of photonic and quasi-optical components in THz systems.

Recent work in CW-THz imaging has highlighted the importance of detailed beam characterization, particularly using coherent detection techniques capable of recovering both amplitude and phase \cite{BenAtar2025}. These methods, based on frequency scanning and fringe analysis, have been successfully used to reconstruct complex THz fields, including those shaped by integrated silicon lenses. Despite their strengths, such approaches require phase-stable interferometric setups, are sensitive to alignment, and are limited by very slow acquisition speeds. 

Only a few other studies have explored the far-field behavior of PIN diode THz photomixers in commercial THz frequency-domain spectroscopy (THz-FDS) systems \cite{Nellen2021,Smith2021}. 
However, these studies relied on one-dimensional (1D) angular scans in the principal E- and H-planes \cite{Nellen2021,Smith2021}. 
While informative, these 1D measurements capture only partial information and may not disclose key features of the full beam profile, including asymmetries, off-axis lobes, and diffraction-induced ring structures. 
These effects become especially pronounced when silicon lenses are integrated with the transmitter, as they introduce complex, frequency-dependent propagation behavior.
In the far-field regime—defined as the region where the electromagnetic field becomes predominantly transverse and angular distribution becomes independent of distance—the spatial beam characteristics can be meaningfully interpreted and compared. This regime begins when:
\begin{equation}
	r \geq \frac{2D^2}{\lambda} = \frac{2D^2\nu}{c}
	\label{eq:1}
\end{equation}
where $r$ is the distance between the transmitter and receiver plane, $D > \lambda/2$ is the effective aperture diameter and $\lambda$ is the free-space wavelength \cite{Balanis2005}. For photomixers with integrated silicon lenses, the high refractive index of silicon ($n_\mathrm{Si} = 3.42$) shortens the wavelength inside the material and extends the near-field region \cite{Smith2021}, complicating the interpretation of beam patterns and potentially contributing to the broadening and structured features observed in prior measurements. Consequently, at frequencies above 0.3 THz, the near-field region can extend beyond the silicon lens, which Smith \textit{et al.} \cite{Smith2021} suggested may explain the beam broadening and additional features observed in their measurements.

In this study, we address these limitations by presenting comprehensive two-dimensional (2D) far-field intensity maps of four nominally identical photomixers across a broad frequency range (0.2–1.5\,THz). Using an incoherent single-pixel detection scheme based on lateral scanning of a broadband Schottky diode receiver, we directly measure the intensity distribution of the emitted THz beam without requiring phase sensitivity. This approach enables robust and high-resolution mapping of key spatial features -- including beam divergence, asymmetries, and frequency-dependent diffraction effects—that are often inaccessible in 1D scans. Our results reveal consistent patterns across devices and frequencies, including clear Airy-like structures resulting from aperture-limited propagation through the silicon lens. These measurements provide new insight into the optical behavior of photomixer-based THz sources and offer valuable benchmarks for transmitter modeling, alignment strategies, and quasi-optical system design.
In addition to qualitative beam mapping, we define and extract a set of robust quantitative metrics -- such as divergence slope, asymmetry, ellipticity, and centroid variability -- that facilitate objective comparison of beam profiles across different antenna designs, fabrication batches, or alignment configurations.

\section{Experimental Work}
\begin{figure}[ht]
\center
\includegraphics[width=\columnwidth]{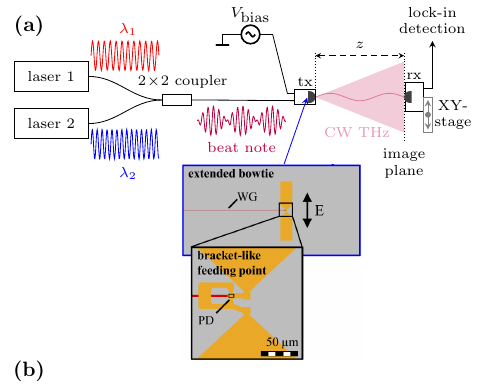}
\includegraphics[trim=0 0 0 10, clip, width=\columnwidth]{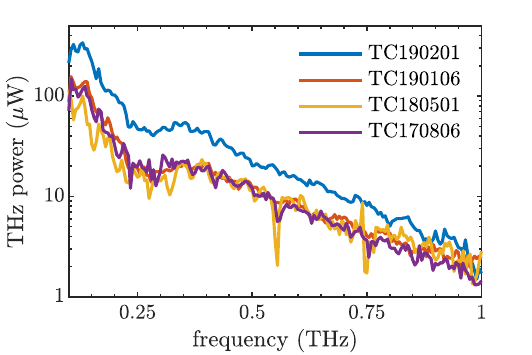}
\caption{
(a) \label{fig:ExperimentalSetup} Illustration of the experimental setup and the bowtie antenna design of the THz photomixer transmitters.
The abbreviations are: tx - photomixer transmitter; rx - Schottky diode receiver; WG - waveguide; PD - PIN photodiode.
The antenna design inset is adapted from Ref. \cite{Nellen2021}, © 2021 Optical Society of America.
(b) \label{fig:THzPowerSpectrum} Power spectra of the four THz photomixer transmitters, adapted from manufacturer data sheets.
}
\end{figure}

The measurements were performed with a TeraScan 1550 system (Toptica Photonics) in combination with an incoherent Schottky diode receiver (Toptica TD-RX-1), which offers a bandwidth of 50\,GHz–1.5\,THz and a noise-equivalent power of 70\,pW/$\sqrt{\mathrm{Hz}}$ at 100\,GHz and 1000\,pW/$\sqrt{\mathrm{Hz}}$ at 1\,THz.
The log-spiral antenna geometry of the Schottky receiver made it insensitive to the polarization of the incident THz radiation.
The receiver was mounted on a motorized XY-stage to perform raster scanning with a 1\,mm step size, enabling full two-dimensional mapping of the far-field emission from THz photomixers. The photomixers themselves were mounted on a Z-stage, allowing the distance to the image plane to be adjusted between 90 and 150\,mm.
The experimental setup is illustrated in Fig. \ref{fig:ExperimentalSetup}(a).

To ensure far-field conditions across all frequencies up to 1.5\,THz, the transmitter-to-receiver distance in our setup exceeds 35\,mm, satisfying the far-field criterion even with the presence of an integrated silicon lens. Previous studies \cite{Nellen2021,Smith2021} have predominantly utilized angular scanning to characterize one-dimensional beam profiles in the E- and H-planes. These measurements were performed by rotating the transmitter relative to the receiver while maintaining polarization alignment. Angular scans are particularly advantageous for divergent beams, as they preserve a constant transmitter-receiver distance and angular accuracy. However, they inherently yield only 1D cross-sections of the beam. Although combining multiple angular scans can reconstruct a 2D radiation pattern, this approach increases experimental complexity and complicates data interpretation. Moreover, even single-plane profiles may produce ambiguous results due to the lack of comprehensive spatial information and overlook beam asymmetries or off-axis features.
In contrast, our study employs a full 2D lateral scan in the far field, enabling a more complete spatial characterization of the THz beam.
While slight distortions are expected due to small variations in propagation distance and the angular sensitivity of the receiver, the resulting 2D profiles provide a significantly more informative picture. For instance, when the transmitter is positioned at $z=$ 100\,mm from a 50\,mm $\times$ 50\,mm image plane, the optical path length to the outermost pixels deviates by only a few percent relative to the center. For each measurement, the transmitter was aligned such that the THz electric field was vertically polarized (E-plane parallel to the $x$-axis) and positioned in front of the Schottky diode at distance $z$.

We characterized four nominally identical InGaAs PIN diode THz photomixers featuring an extended bowtie antenna geometry (See inset in Fig. \ref{fig:ExperimentalSetup}(a)) with an effective aperture diameter of approximately $D\approx 1$\,mm.
The power spectra of these devices, adapted from test reports provided by the Fraunhofer Heinrich Hertz Institute (HHI), are shown in Fig. \ref{fig:THzPowerSpectrum}(b).
All four antennas perform within specified parameters; however, one device (indicated by the blue line) clearly outperforms the others in terms of output power. This particular unit (TC190201) originates from a newer production batch. Despite this, all four photomixers are based on the same chip design, and according to the manufacturer, no changes have been made to the fabrication or assembly processes.
\footnote{According to HHI, based on private correspondence with Dr. Anselm Deninger, Director Technical Sales Support, Toptica Photonics AG.}

\section{Results and Discussion}
\begin{figure}
\centering
\includegraphics[width=\columnwidth]{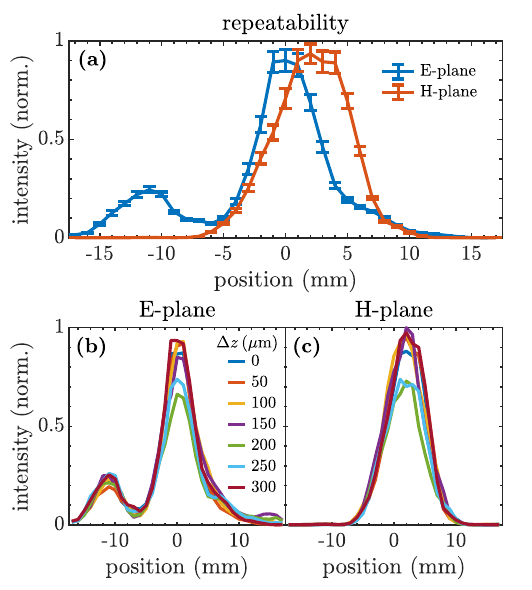}
\caption{Measurement repeatability test at $\nu = 0.5$\,THz using the TC190201 antenna (a), and evaluation of the influence of standing waves in the setup (b) and (c).}
\label{fig:THzbeamPattern-repeatability}
\end{figure}

First, we evaluated the measurement repeatability of our setup. Ten successive measurements were performed on a single antenna at $z = 145$\,mm, and the resulting means and standard deviations for the E- and H-planes are shown in Fig. \ref{fig:THzbeamPattern-repeatability}(a). Overall, the repeatability is good, with the largest variance occurring near the main peak. Moreover, the general beam profiles in both the E- and H-planes are in good agreement with previously published results \cite{Smith2021,Nellen2021}.

Standing waves between the transmitter and receiver can affect measurements when using CW THz radiation, and must therefore be considered. To assess their influence, we scanned the receiver position over a 300\,$\mu$m interval in steps of 50\,$\mu$m between consecutive beam profile measurements at 0.5 THz. This interval corresponds to approximately $\lambda/2$. The results for the E- and H-planes are shown in Fig. \ref{fig:THzbeamPattern-repeatability}(b) and (c), respectively. A noticeable decrease in the peak amplitude is observed at $\Delta z = 200$ and 250\,$\mu$m, which may indicate the presence of standing wave effects when the receiver is positioned directly in front of the transmitter.

\begin{figure}
\centering
\includegraphics[trim=0 50 0 15, clip, width=.85\columnwidth]{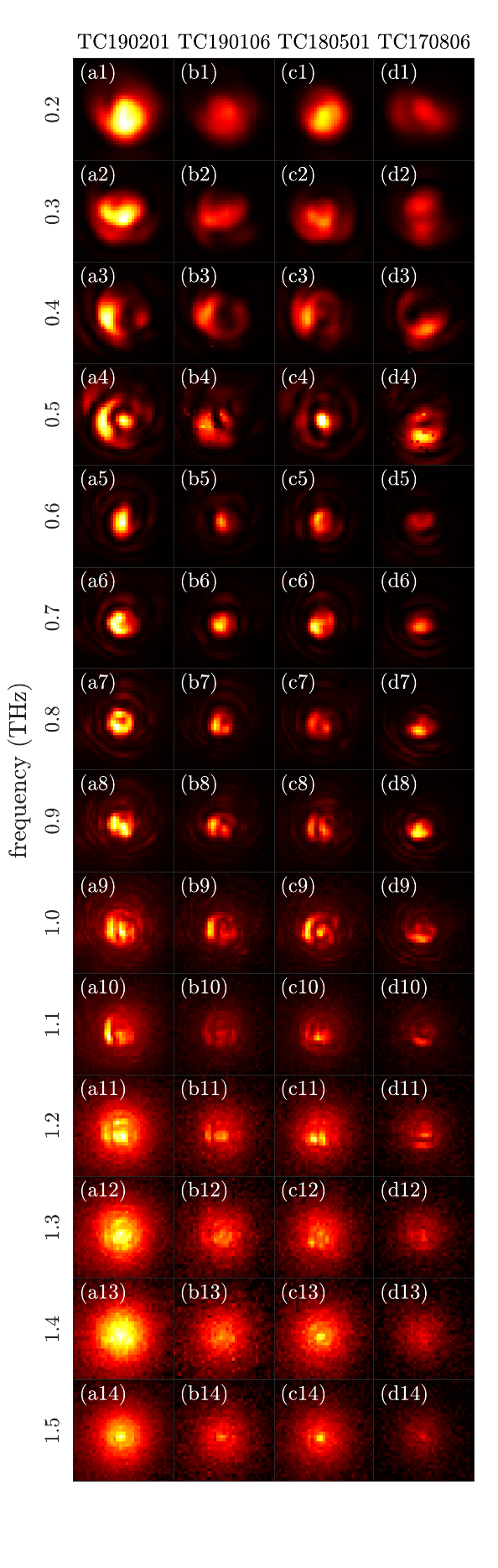}
\caption{Radiation patterns of the (a) TC190201, (b) TC190106, (c) TC180501, and (d) TC170806 antennas at frequencies from 0.2 to 1.5\,THz. Each map covers a 35\,mm\,$\times$\,35\,mm area in 1\,mm increments and is measured at a distance $z = 90$\,mm in front of the antennas.}
\label{fig:THzRadiationPattern}
\end{figure}

\begin{figure}[ht]
\center
\includegraphics[width=\columnwidth]{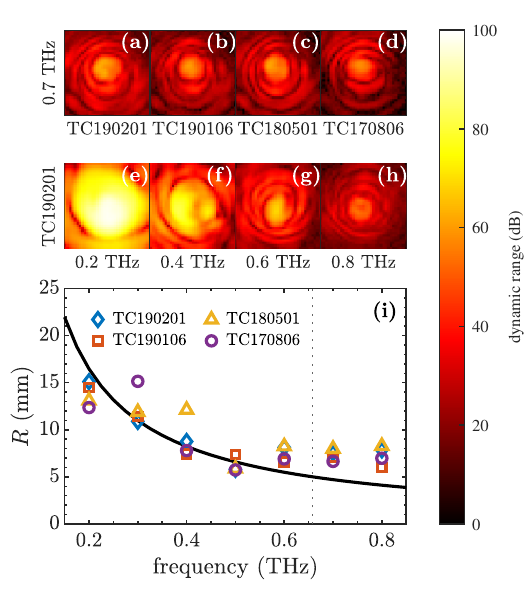}
\caption{
Dynamic range of the emission patterns (35\,mm$\times$35\,mm in 1\,mm increments) measured at $z = 90$\,mm:
(a)-(d) \label{fig:DynamicRange700GHz}  Of all four transmitters as measured at $\nu=$0.7 THz.
(e)-(h) \label{fig:DynamicRange-TC190201}  Of the TC190201 antenna at frequencies $\nu=$ 0.2, 0.4, 0.6 and 0.8\,THz.
And (i) estimated Airy disk radius as a function of frequency of the four THz transmitters. The solid black curve is the theoretical Airy disk radius assuming a 10-mm aperture radius.
The dotted vertical line represent the maximum frequency, for which at least five measurements across the Airy disk can be obtained using 1 mm increments.
}
\label{fig:THzAiryDisk}
\end{figure}

The radiation patterns were measured over a 35\,mm\,$\times$\,35\,mm area in 1\,mm increments for frequencies ranging from 0.2\,THz to 1.5\,THz, with the receiver positioned at $z = 90$\,mm in front of the antennas. The resulting beam profiles for the four transmitters are presented in Fig. \ref{fig:THzRadiationPattern}. For consistency, the color scale in each map is normalized to the output of the TC190201 antenna at the corresponding frequency.
At a first glance, the beam diameter decreases from 0.2\,THz, reaching a minimum around 0.6$\sim$0.9\,THz, before increasing again up to 1.5\,THz. Additionally, the radiation patterns exhibit irregularities particularly around 0.4$\sim$0.5\,THz and 1$\sim$1.2\,THz. At higher frequencies, the patterns appear increasingly blurred, which can be attributed to a reduced signal-to-noise ratio and the fixed 1-mm spatial resolution.
The observed asymmetries in the E- and H-plane profiles have previously been attributed by Nellen \emph{et al.}\,\cite{Nellen2021} to the asymmetric feed point structure of the antenna, as shown in Fig. \ref{fig:ExperimentalSetup}(a). Notably, the dimensions of the bracket structure correspond to $\lambda/4$ at 0.35\,THz, and the extended bowtie geometry has been proposed to act as a broad dipole, generating frequency-dependent side lobes. 
A closer examination of the beam profiles in the 0.5–1\,THz range reveals the presence of weaker rings surrounding the main lobe.
In Fig. \ref{fig:DynamicRange700GHz}(a)-(d), the dynamic range at 0.7\,THz is calculated for each antenna to highlight the radiation pattern.
In all cases, an Airy-like pattern is observed — characterized by a bright central spot (Airy disk) surrounded by concentric bright rings — arising from diffraction through a circular aperture.
Notably, the Airy disk appears approximately the same size across all antennas.
Fig. \ref{fig:DynamicRange-TC190201}(e)-(h) shows the dynamic range maps of the TC190201 antenna across the 0.2–0.8\,THz frequency range.
Here, the diameters of both the central spot and the concentric rings are seen to decrease with increasing frequency.
This behavior may be attributed to diffraction of the THz waves by the circular aperture of the integrated silicon lens.
The minima of an Airy pattern observed far from the aperture occur at angles given by
\begin{equation}
\sin\theta \approx m \frac{\lambda}{d},
\label{eq:2}
\end{equation}
measured relative to the direction of the incident light, where $d$ is the diameter of the aperture, and $m = 1.220,\ 2.233,\ 3.238$ corresponds to the first three diffraction minima from a circular aperture \cite{Hecht2002}.
Accordingly, the diameter of the Airy disk is expected to be inversely proportional to the frequency $\nu = c / \lambda$.

To estimate the Airy disk radius from the measured beam intensity profiles, we fit the normalized intensity data along a single spatial dimension (e.g., $x$, $y$, diagonal, and anti-diagonal) to a theoretical Airy pattern given by
\begin{equation}
    I(r) = \left[ \frac{2 J_1\left( \alpha\,r \right)}{\alpha\,r} \right]^2,
\end{equation}
where $r$ is the radial coordinate, $J_1$ is the first-order Bessel function of the first kind, $\alpha = \pi d \nu /c z$ with $c$ the speed of light, and $z$ the propagation distance from the circular aperture to the detection plane \cite{Hecht2002}. 
In practice, we fit the experimental data to this function using $\alpha$ as a free parameter.
The first minimum of the Airy pattern, which defines the Airy disk radius $R$, occurs where $J_1(\alpha\, r)$ first vanishes. 
Since the first zero of $J_1(x)$ occurs at $x \approx 3.8317$, we obtain the relation $R = 3.8317/\alpha$.
This constant thus directly links the fitted parameter $\alpha$ to the physical Airy disk radius, providing a reliable means to extract the diffraction-limited spot sizes from the experimental data.
To improve the robustness of the estimate, we perform the fit along the $x$, $y$, diagonal, and anti-diagonal directions, and then calculate the mean value of the resulting fits.
The results for the four THz transmitters between 0.2 and 0.8\,THz are shown in Fig. \ref{fig:THzAiryDisk}(i).
The solid black curve represents the theoretical Airy disk radius
\begin{equation}
    R(\nu) = \frac{3.8317\, c\,z}{\pi d \nu},
\end{equation}
derived from the definition of the scaling parameter $\alpha$, assuming a circular aperture diameter $d=10$\,mm, corresponding to that of the silicon lens.
A good agreement is observed between the theoretical and the measured radii.
Above 0.6\,THz, however, the measured Airy disk radii exceed the theoretical predictions for all four transmitters.
This discrepancy is likely due to the limited spatial resolution of the measurement system, effectively imposing the 1-mm increment size.
To determine the maximum frequency at which the Airy disk can be reliably resolved under this constraint and a propagation distance of $z = 90$\,mm, we require at least five sampling points per Airy disk radius.
Solving the condition $R(\nu)\,/\,1\,\mathrm{mm} \geq 5$ yields a maximum resolvable frequency $\nu_\mathrm{max} \approx 0.66$\,THz (indicated by the dotted vertical line), consistent with the observed deviation.
We thus conclude that the radiated THz waves are diffracted by the effective aperture of the entire device, resulting in an Airy pattern.
When observed solely along the E- and H-planes, this pattern may be misinterpreted as side lobes.

\begin{figure}[ht]
\center
\includegraphics[width=\columnwidth]{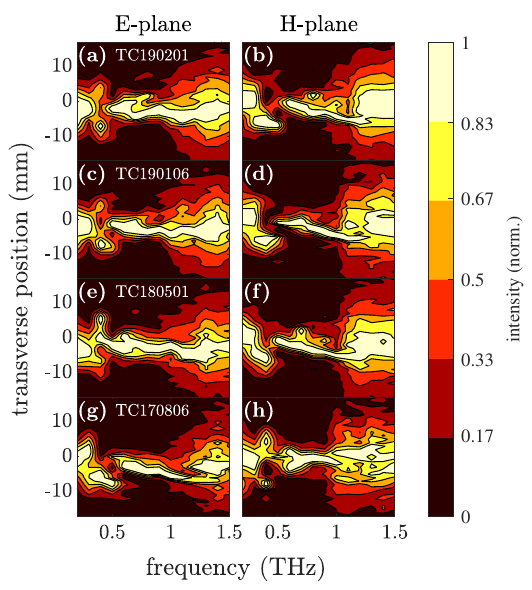}
\caption{
E-plane (left) and H-plane (right) spatial THz beam profiles as a function of frequency for the four norminally identical photomixer transmitters. At each frequency, the intensity is normalized to its maximum value to highlight spatial distribution and beam shape. The plots reveal device-to-device differences in beam divergence, symmetry, and frequency-dependent steering behavior.
}
\label{fig:spatial_distribution}
\end{figure}

Fig. \ref{fig:spatial_distribution} presents the frequency-dependent spatial beam profiles in the E- and H-plane for the four transmitters.
The intensity at each frequency is normalized to its maximum to emphasize the relative spatial distribution and beam shape. 
Despite identical design and fabrication, clear differences are observed in beam divergence, symmetry, and spatial stability across the transmitters.
In particular, some devices show broader and more diverging beams at low frequencies, while others exhibit off-axis emission or asymmetric lobe structures. 
These qualitative variations highlight the importance of characterizing each emitter individually, as small differences in antenna fabrication, lens alignment, or packaging can lead to appreciable changes in beam characteristics. 
To quantify these observations, we extracted beam centroid positions, angular emission direction, root-mean-square (RMS) beam widths, divergence slopes, and symmetry metrics across the full frequency range (See Table \ref{tab:metrics}. 
The following sections present these metrics and evaluate emitter-to-emitter variability, providing insight into reproducibility and alignment sensitivity in commercial THz systems.

\begin{table*}[ht]
\centering
\caption{Summary of beam characteristics for the four nominally identical THz photomixer transmitters.}
\label{tab:metrics}
\begin{tabular}{@{}lcccc@{}}
\toprule
         & TC190201 & TC190106 & TC180501 & TC170806 \\
\midrule
Max angle deviation (E) [deg]       & 1.23  & 1.00  & 1.07  & 2.50 \\
Max angle deviation (H) [deg]       & 2.99  & 3.05  & 3.26  & 0.60 \\[.125cm]
Mean RMS width (E) [mm]             & 6.1   & 6.1   & 6.0   & 5.8 \\
Mean RMS width (H) [mm]             & 6.1   & 6.2   & 6.2   & 6.0 \\[.125cm]
Divergence slope (E) [mm/THz]       & 2.48  & 2.62  & 2.77  & 2.31 \\
Divergence slope (H) [mm/THz]       & 2.49  & 2.48  & 2.88  & 2.36 \\[.125cm]
Mean asymmetry (E)                  & 0.3633 & 0.3770 & 0.4045 & 0.5003 \\
Mean asymmetry (H)                  & 0.2912 & 0.3068 & 0.3131 & 0.2105 \\[.125cm]
Mean ellipticity (E/H)              & 0.9915 & 0.9803 & 0.9747 & 0.9858 \\
\bottomrule
\end{tabular}
\end{table*}

\subsection{Beam Centroid and Steering}
To quantitatively assess beam pointing behavior, we extracted the spatial centroid position at each frequency for all transmitters in both the E- and H-planes.
The resulting centroid trajectories reveal frequency-dependent steering effects, particularly in the H-plane, where some transmitters exhibited angular deviations up to -3.3$^\circ$, consistent with Fig. \ref{fig:spatial_distribution}. 
These angles were computed from the centroid shift assuming a propagation distance of $z=90$\,mm.
To account for transmitter alignment offsets, all emission angles were calculated relative to the centroid position at 0.2 THz. 
This ensures the observed angular trends reflect frequency-dependent steering rather than absolute misalignment of the transmitters.
The angular variation may originate from off-center illumination of the effective aperture, which can shift the central lobe of the Airy pattern. 
The standard deviation of emission angle across devices peaked near 0.4–0.6\,THz, but stabilized at higher frequencies where the central Airy pattern becomes more delicate and could not be reliably resolved due to the limited spatial resolution of the measurement system.

\subsection{Beam Widths and Divergence}
To characterize the frequency-dependent beam divergence of the transmitters, we analyzed the RMS beam widths (see Table \ref{tab:metrics}). 
Linear fits of RMS width versus frequency yielded divergence slopes ranging from 2.3 to 2.8\,mm/THz (E-plane) and 2.4 to 2.9\,mm/THz (H-plane) across the transmitters. 
These values reflect a consistent, moderate increase in beam width with frequency.
When converted to FWHM-equivalent values (assuming Gaussian scaling), this corresponds to 5.4$\sim$6.8\,mm/THz.
For comparison, the vendor-reported full-angle FWHM divergence of 12$^\circ$ at 0.2\,THz and 15$^\circ$ from 0.5 to 1.0\,THz correspond to an approximate divergence rate of $\sim$\,8\,mm/THz at $z=90$\, mm.
This difference may result from our use of RMS-based beam radii or reflect device-to-device variation in alignment or illumination of the aperture. Moreover, the vendor-reported FWHM divergence may be an upper limit.
A slight increase in the RMS width is observed above 1\,THz, which likely results from spatial under-sampling beyond the resolvable Airy disk limit ($\sim$0.66\,THz) rather than a true increase in beam divergence.

\subsection{Beam Shape and Symmetry}
Clear asymmetries can be seen at certain frequencies in both the E- and H-planes among all transmitters in Fig. \ref{fig:spatial_distribution}. 
We quantify the beam shape and symmetry by using two complementary metrics: asymmetry, which measures the lateral imbalance of intensity about the beam center, and ellipticity, which quantifies the overall geometric elongation of the beam in the E- and H-planes.
Together, asymmetry and ellipticity offer a detailed view of beam shape stability, helping to distinguish between lateral imbalance and global shape distortion -- both of which are important for evaluating transmitter reproducibility and far-field beam quality.

Beam asymmetry was calculated as the normalized intensity imbalance between the two sides of the beam relative to its center, defined as $A=\lvert I_\mathrm{left}-I_\mathrm{right}\rvert/(I_\mathrm{left}+I_\mathrm{right})$, where $I_\mathrm{left}$ and $I_\mathrm{right}$ are the integrated intensities on either side of the centroid position.
This metric revealed significant device-to-device variation, especially at lower frequencies.
In the E-plane, TC170806 showed the most pronounced asymmetry, reaching 0.82 at 0.4\,THz and maintaining a mean asymmetry of 0.52 across all frequencies. 
In contrast, the other transmitters exhibited more balanced profiles, with mean values between 0.38 and 0.40. 
The H-plane asymmetries were generally lower, averaging 0.25–0.34. 
At higher frequencies ($\geq$1.0\,THz), asymmetry values dropped below 0.25 for all transmitters, consistent with the symmetric beam shape due to the reduced spatial resolution, as seen in both Fig. \ref{fig:THzRadiationPattern} and Fig. \ref{fig:spatial_distribution}.

Ellipticity was defined as the ratio of the RMS beam widths in the E- and H-planes (i.e. $\varepsilon=w_\mathrm{E}/w_\mathrm{H}$), with $\varepsilon=1$ indicating a circular beam and deviations reflecting anisotropic beam shapes.
Mean ellipticity values ranged from 0.9747 to 0.9915, indicating that the beams were nearly circular on average.
However, frequency-dependent deviations were observed, particularly between 0.3–0.6\,THz, where $\varepsilon$ reached as high as 1.30 (E-plane broader) and as low as 0.68 (H-plane broader).
These shape distortions likely result from subtle misalignments and off-axis aperture illumination.

\subsection{Transmitter Variability}
To assess the consistency between nominally identical transmitters, we quantified device-to-device variability by computing the standard deviation of the beam centroid positions and the RMS beam widths across all four transmitters at each frequency.
These metrics capture the extent to which each photomixer deviates from the average beam shape and pointing behavior, providing insight into the reproducibility of commercially packaged THz sources.
We found that variability peaks in the 0.4–0.6\,THz range, where standard deviations in both centroid position and beam width are largest. 
This frequency range coincides with the transition region where the beam narrows, diffraction rings begin to emerge, and the feed structure becomes electrically significant (approaching $\lambda/4$) \cite{Nellen2021}. 
The enhanced sensitivity in this region suggests that small differences in feed geometry, lens alignment, or emitter placement can lead to appreciable changes in beam shape and pointing.
At lower frequencies (0.2–0.3\,THz), the relative variability is less pronounced due to the much broader Airy disk and reduced sensitivity to small differences.
In contrast, at higher frequencies (>0.9\,THz), variability decreases again, likely for two reasons:
(i) the Airy disk becomes smaller, reducing the spatial extent of measurable deviations, and
(ii) the 1\,mm scan resolution limits the ability to resolve fine-scale differences in beam shape or centroid position.
This resolution-induced smoothing may partially obscure real variations but still reflects the practical limits of reproducibility for high-frequency THz beams in commercial CW measurement setups.

These findings underscore that even in nominally identical commercial devices, sub-wavelength differences in alignment or structure can manifest as measurable beam divergence, skew, and asymmetry. 
For applications requiring tight spatial control this variability highlights the importance of per-emitter characterization and alignment to ensure optimal system performance.

\subsection{Comparison to Prior Work}
As mentioned, prior beam profile studies of CW THz photomixers have relied on 1D angular scans to characterize the E- and H-plane emission patterns.
In this section, we critically compare our 2D far-field mapping approach and findings with these prior efforts by Nellen et al. \cite{Nellen2021} and Smith et al. \cite{Smith2021}.

Nellen et al. investigated the far-field beam patterns of custom-designed photomixers incorporating three distinct bowtie antenna variants: an extended bowtie with a bracket-like feeding structure, an extended bowtie with a centered feed, and a triangular bowtie with a merged feed point. 
Using 1D angular scans, they demonstrated that antenna feed geometry has a pronounced effect on beam tilt, symmetry, and the emergence of side lobes. 
In particular, the merged feed structure produced narrower and more symmetric beams, whereas the bracket-like feed was associated with increased asymmetry and beam steering.
In our study, although all transmitters share a nominally identical chip design (featuring an extended bowtie with a bracket feed), we observe notable device-to-device variability in beam symmetry, centroid position, and angular emission direction (see Table \ref{tab:metrics}). 
This variation likely reflects subtle differences in packaging and lens alignment, which can significantly influence far-field beam characteristics.
While our analysis identifies optical diffraction as a key factor influencing the beam pattern, it likely acts through an interplay with alignment sensitivity and feed geometry, suggesting a multifaceted origin of the observed asymmetries.
In particular, the most pronounced asymmetries and angular deviations occur between 0.3 and 0.6\,THz -- precisely the frequency range where the electrical length of the feed approaches $\lambda/4$ and may strongly perturb the near-field radiation \cite{Nellen2021}.
Moreover, our asymmetry metrics show device-specific trends that may not be fully explained by misalignment alone. 
For instance, transmitter TC170806 exhibits higher mean asymmetry and larger centroid shifts than the other transmitters.
These observations suggest that even among nominally identical devices, microscopic variations in the feed structure or bonding geometry may contribute to beam distortions, particularly in the mid-THz range.

Smith et al. carried out 1D angular beam profile measurements of a single commercial silicon-lens-integrated photomixer (comparable to the transmitters in this study) over a wide frequency range (0.1–1\,THz)
Their study highlighted increasing beam broadening and profile skewness above 0.4\,THz, attributing these effects to substrate radiation, internal reflections, and feed asymmetries. 
They also emphasized the potential role of the near-field/far-field transition and advocated for more rigorous metrological analysis of THz beam profiles.
Our results complement and refine these interpretations.
Like Smith et al., we observe increased asymmetry and beam distortion in the 0.3–0.6\,THz range.
However, our 2D beam profiles and Airy-pattern fitting demonstrate that the dominant beam structure across all devices and frequencies is well described by diffraction through a circular aperture -- namely, the silicon lens.
The asymmetric and skewed features noted in Smith’s study can thus be reinterpreted as perturbations of an underlying Airy envelope, arising from off-axis illumination of the photomixer, lateral lens displacement, or lens-edge interactions, rather than solely from electromagnetic feed effects.
Furthermore, our results show that beam symmetry improves above $\sim$1.0\,THz, coinciding with the decreasing Airy disk radius and limited spatial resolution of our scan system. 
This behavior contrasts with Smith’s attribution of high-frequency distortion to internal reflections or feeding structure artifacts and suggests that some of their observed broadening at high frequency may reflect measurement limitations rather than physical emission effects.

\section{Conclusion}
We performed a comprehensive two-dimensional far-field characterization of four nominally identical, commercially available PIN diode THz photomixers featuring extended bowtie antenna structures.
Despite their identical chip designs and packaging, we observed device-to-device variations in beam width, angular emission direction, and asymmetry.

Across all devices and frequencies, a consistent beam structure emerged: a dominant central lobe surrounded by weak concentric rings.
These Airy-like patterns are well explained by diffraction through the circular aperture defined by the silicon hyperhemispherical lenses, as confirmed by quantitative fits to the theoretical Airy function.
Our analysis indicates that this aperture-induced diffraction is a key factor shaping the far-field beam envelope, with additional contributions from misalignment and possible microstructural variation in feed geometry.

To evaluate emitter reproducibility, we extracted beam centroids, divergence slopes, angular pointing deviations, and asymmetry metrics.
While key beam parameters such as divergence slope (5.4–6.8\,mm/THz at FWHM) and angular deviations (up to $\sim$3.3$^\circ$) varied among devices, the main beam shape remained robustly governed by diffraction, underscoring the dominant role of the lens aperture in determining far-field emission.

These findings have practical implications for THz imaging, spectroscopy, and communication systems where precise beam shaping and alignment are critical. 
Even small deviations in lens alignment or optical coupling can shift the beam center, alter divergence, and impact spatial overlap with downstream optics. 
This highlights the importance of emitter-specific beam profiling, especially in multi-emitter or standoff setups requiring high spatial precision.

In principle, such diffraction-induced effects could be mitigated through monolithic integration of micro-structured lenses directly onto the antenna substrate \cite{Brincker2016,Sondergaard2020}. 
Nonetheless, even with conventional silicon lens packaging, our results provide valuable benchmarks for system design and support continued development of advanced THz emitter architectures.

\section*{Acknowledgment} 
We gratefully acknowledge Dr. Anselm Deninger, Director of Technical Sales Support at TOPTICA Photonics AG, for his insightful discussions, for providing valuable details on the photomixer antenna design, and for sharing the calibrated power spectra of the photomixers.

\section*{Funding} This work was supported by the Innovation Fund Denmark Grand Solutions program (grant no. IFD-7076-00017B).

\section*{Disclosures} The authors declare no conflicts of interest.

 \bibliographystyle{elsarticle-num} 
 \bibliography{photomixer_emission_patterns}





\end{document}